\let\accentvec\vec
\documentclass[runningheads,a4paper]{llncs}

\let\vec\accentvec
\usepackage{amsmath,amssymb}
\usepackage{graphicx}
\usepackage{hyperref}
\usepackage{url}
\usepackage{listings}
\usepackage{enumitem}

\usepackage{xcolor}
\definecolor{darkgreen}{rgb}{0.0 0.5 0.0}
\definecolor{whitegreen}{rgb}{0.0 0.75 0.0}
\definecolor{whiteblue}{rgb}{0.0 0.0 1.2}
\definecolor{darkred}{rgb}{0.8 0.0 0.0}
\definecolor{dkblue}{rgb}{0,0.1,0.5}
\definecolor{lightblue}{rgb}{0,0.5,0.5}
\definecolor{dkgreen}{rgb}{0,0.4,0}
\definecolor{dk2green}{rgb}{0.4,0,0}
\definecolor{dkviolet}{rgb}{0.6,0,0.8}
\definecolor{dkpink}{rgb}{0.8,0,0.9}

\newcommand{\R}{\mathbb{R}}					

\newcommand{\fpop}{f_{\text{pop}}}
\newcommand{\epspop}{\epsilon_{\text{pop}}}

\newcommand{\K}{\mathbf{K}}

\DeclareMathOperator{\parab}{par}

\newcommand{\xb}{\mathbf{x}} 
\newcommand{\xopt}[7]{\xb_{#1} := (#2, #3, #4, #5, #6, #7)}

\newcommand{\code}[1]{\lstinline{#1}}

\newcommand{\nlcertify}{\mathtt{NLCertify}}
\newcommand{\coq}{\textsc{Coq}}
\newcommand{\sdpa}{\text{\sc Sdpa}}

\newcommand{\ssreflect}{\text{\sc Ssreflect}}

\newcommand{\hol}{\text{\sc Hol-light}}
\newcommand{\ocaml}{\text{\sc OCaml}}
\newcommand{\micromega}{\mathtt{micromega}}
\newcommand{\coqinterval}{\mathtt{interval}}

\newcommand{\sollya}{\mathtt{Sollya}}

\usepackage{amsfonts} 
\usepackage{tikz, pgfplots}
\def\samplepar{10}
\def\afst{2.5}
\def\asnd{0.5}
\def\athd{1.5}
\def\lo{0.13}
\def\up{4}
\def\sampleatn{30}

\newcommand{\keywords}[1]{\par\addvspace\baselineskip
\noindent\keywordname\enspace\ignorespaces#1}

\begin{document}

\mainmatter

\title{{\texttt{NLCertify}}: A Tool for Formal Nonlinear Optimization}  
\titlerunning{$\nlcertify$: A Tool for Formal Nonlinear Optimization} 
\author{Victor Magron\inst{1}}
\authorrunning{Victor Magron}
\institute{
~$^{1}$LAAS-CNRS, 7 avenue du colonel Roche, F-31400 Toulouse, France\\
\email{magron@laas.fr},\\ 
\texttt{http://homepages.laas.fr/vmagron}
}
\maketitle

\begin{abstract}
$\nlcertify$ is a software package for handling formal certification of nonlinear inequalities involving transcendental multivariate functions. The tool exploits sparse semialgebraic optimization techniques with approximation methods for transcendental functions, as well as formal features. Given a box and a transcendental multivariate function as input, $\nlcertify$ provides $\ocaml$ libraries that produce nonnegativity certificates for the function over the box, which can be ultimately proved correct inside the $\coq$ proof assistant. 
\end{abstract}

\section{Introduction}
A variety of tools for solving nonlinear systems are being adapted for the field of formal reasoning. One way to import the technology available inside an {\em informal} tool is the {\em skeptical} approach: the tool yields a form of certificate which can be verified on the {\em formal} side, i.e. inside a theoretical prover such as $\coq$~\cite{CoqProofAssistant}. A recent illustration~\cite{armand:hal-00639130} is the integration of the computational features of SAT/SMT solvers in $\coq$.
The $\nlcertify$\footnote{The source code is available at \url{https://forge.ocamlcore.org/frs/?group_id=351}.  See also the documentation at \url{http://nl-certify.forge.ocamlcore.org}.} tool has informal nonlinear optimization features and enables formal verification in a skeptical way. An ambitious application is to automatically verify real numbers inequalities occurring  by thousands in Thomas Hales' proof of Kepler's conjecture~\cite{halesannmath}. 
In the present article, nonlinear functions include polynomials, semialgebraic functions obtained by composition of polynomials with some basic operations (including the square root, $\sup$, $\inf$, $+$, $\times$, $-$, $/$, etc.) and composition of semialgebraic functions with transcendental functions ($\arctan$, $\cos$, $\exp$, etc.) or basic operations.

Polynomial inequalities over a finite set of polynomial constraints can be certified using a hierarchy of sums of squares (SOS) relaxations~\cite{Lasserre01moments}. Several variants of these relaxations are implemented in some {\sc MATLAB} toolboxes: {\sc GloptiPoly} 3~\cite{henrion:hal-00172442} solves the Generalized Problem of Moments 
while {\sc SparsePOP} takes sparsity into account~\cite{DBLP:journals/toms/WakiKKMS08}, {\sc YALMIP}~\cite{YALMIP} is a high-level parser for nonlinear optimization problems and has a built-in module for SOS calculations. These toolboxes rely on external SOS solvers for solving the relaxations (e.g. 
$\sdpa$~\cite{YaFuNaNaFuKoGo:10})
.
However, the validity of the bounds obtained with these numerical tools can be compromise, due to the rounding error of the SOS solver. The tool\footnote{Available from the pages~\url{http://bit.ly/fBNLhR} and~\url{bit.ly/gPXNF8}} mentioned in~\cite{Monniaux_Corbineau_ITP2011} allows to handle some degenerate situations.
For a more general class of problems (when the functions are not restricted to polynomials), one can combine SOS software with frameworks that approximate transcendental functions. $\sollya$~\cite{ChevillardJoldesLauter2010} returns safe tight bounds for the approximation error obtained when computing minimax estimators of nonlinear univariate functions.

On the formal side, recent efforts have been done to verify nonlinear inequalities with theorem provers. A tool\footnote{\url{http://flyspeck.googlecode.com/files/FormalVerifier.zip}} in $\hol$  combines formal interval arithmetic computation and quadratic Taylor approximation~\cite{DBLP:journals/corr/abs-1301-1702}.
The features of the {\sc MetiTarski}~\cite{Akbarpour:2010:MAT:1731476.1731498} theorem prover include continued fractions expansions of univariate transcendental functions such as $\log$, $\arctan$, etc. {\sc PVS} incorporates nonlinear optimization libraries relying on Bernstein polynomial approximation~\cite{PVS2013Bernstein}. The $\coqinterval$\footnote{\url{https://www.lri.fr/~melquion/soft/coq-interval/}} tactic can assert the validity of interval enclosures of nonlinear functions over a finite set of box constraints inside $\coq$.  The $\micromega$ tactic returns emptiness certificates for basic semialgebraic sets~\cite{Besson06micromega}.

One specific challenge of the field of formal nonlinear optimization is to develop adaptive techniques to produce certificates with a reduced complexity.
$\nlcertify$ provides efficient informal libraries by implementing the nonlinear maxplus method~\cite{mapr14}, which combines sparse SOS relaxations with maxplus quadratic approximation. In addition, the tool offers a secure certification framework for the bounds obtained with these semialgebraic relaxations~\cite{jfr14}. These various features are placed in a unified framework extending to about 15000 lines of $\ocaml$ code and 3600 lines of $\coq$ code. The $\nlcertify$ package can solve successfully non-trivial inequalities from the Flyspeck project (essentially tight inequalities, involving both semialgebraic and transcendental expressions of 6 to 12 variables) as well as significant global optimization benchmarks. The running tests for the verification of polynomial inequalities (Section~\ref{sec:pop}) and transcendental inequalities (Section~\ref{sec:trans}) are performed on Intel Core i5 CPU ($2.40\, $GHz)\footnote{With $\ocaml$ $4.01.0$, $\coq$ 8.4pl2, $\ssreflect$ 1.4, $\sdpa$ 7.3.6 and $\sollya$ 3.0}.
\section{Certified Polynomial Optimization}
\label{sec:pop}

One particular problem among certification of nonlinear problems is to verify the inequality $\forall \xb \in \K, \fpop (\xb) \geq 0$, where $\fpop$ is an $n$-variate positive polynomial, $\K := \{\xb\in \R^n : g_1 (\xb) \geq 0, \dots, g_m(\xb) \geq 0\}$ is a semialgebraic set obtained with polynomials $g_1, \dots, g_m$.
One way to convexify this polynomial optimization problem is to find sums of squares of polynomials $\sigma_0, \sigma_1, \dots, \sigma_m$ satisfying 
$\fpop (\xb) = \sigma_0 + \sum_{j = 1}^m \sigma_j(\xb) g_j(\xb)$ and $\deg \sigma_0 \leq 2 k, \deg (\sigma_1 g_1 )  \dots, \deg (\sigma_m g_m) \leq 2 k$, for a fixed positive integer $k$ (called the {\em relaxation} order). When $k$ increases, one obtains progressively stronger relaxations. In this way, it is always possible to certify the inequality $\forall \xb \in \K, \fpop (\xb) \geq 0$ for a sufficiently large order (under certain assumptions~\cite{Lasserre01moments}).
These relaxations are implemented in $\nlcertify$ and numerically solved with the SOS solver $\sdpa$. 
The tool performs a rational extraction from the SOS solver output with the {\sc Lacaml}\footnote{Linear Algebra with $\ocaml$, this library implements {\sc Blas}/{\sc Lapack} routines} library. Then the corresponding remainder $\epspop$ (the difference between the objective polynomial $\fpop$ and the SOS representation) can be bounded on a box which contains $\K$. More details can be found in~\cite{jfr14}.
\begin{example} (caprasse)
\label{ex:caprasse}
Here, we consider the degree 4 polynomial inequality $\forall \xb \in [-0.5, 0.5]^4,  -x_1 x_3^3 + 4 x_2 x_3^2 x_4 + 4 x_1 x_3 x_4^2 + 2 x_2 x_4^3 + 4 x_1 x_3 + 4 x_3^2 - 10 x_2 x_4 - 10 x_4^2 + 5.1801 \geq 0 $. 
The inequality is scaled on $[0, 1]^4$ with the solver option \code{scale_pol = true} and one adds the redundant constraints $x_1^2 \leq 1, \dots, x_4^2 \leq 1$ by setting \code{bound_squares_variables = true}. The inequality can be solved numerically at the second SOS relaxation order (\code{relax_order = 2}).  The correctness of the SOS representation is verified inside $\coq$ (via the reflexive tactic \code{ring}) by setting \code{check_certif_coq  = true}. Then the execution of $\nlcertify$ returns the following output:
\lstset{language=NLcertify}
\scriptsize
\begin{lstlisting}
% ./nlcertify caprasse
Proving that - x1 * x3 * x3 * x3 + 4 * x1 * x3 * x4 * x4 + 4 * x1 * x3 + 4 *
x2 *x3 * x3 * x4 + 2 * x2 * x4 * x4 * x4 - 10 * x2 * x4 + 4 * x3 * x3 - 10 * 
x4 * x4 + 5.1801 >= 0 over the box [(-0.5, 0.5); (-0.5, 0.5); (-0.5, 0.5); (-0.5, 0.5)]
...
Computing lower bound ...
 SOS numerical computation in 0.045087 secs
Proving non-negativity inside Coq
     = true
     : bool
Finished transaction in  1. secs (0.813333u,0.s)
Lower Bound with SOS of degree at most 4 = 0.0000021671
...
0.0000021642 >= 0.0000000000
Inequality caprasse verified
\end{lstlisting}
\normalsize
Here, the caprasse inequality is formally proved in less than 1 second which is 8 times faster than the verification procedure in $\hol$ with the framework described in~\cite{DBLP:journals/corr/abs-1301-1702} and 10 times faster than the tool based on Bernstein approximation in {\sc PVS}~\cite{PVS2013Bernstein}.
\end{example}
\section{Certificates for Nonlinear Transcendental Inequalities}
\label{sec:trans}
Now, we consider a more general goal $\forall \xb \in \K, f(\xb) \geq 0$, where $f$ is an $n$-variate transcendental function and $\K \subset \R^n$ is a box.
$\nlcertify$ implements the nonlinear maxplus method~\cite{mapr14}, which can be summarized as follows. The tool builds first the abstract syntax tree $t$ of $f$ (see Figure~\ref{fig:ast_9922} for an illustration).
The leaves of $t$ are semialgebraic functions. The other nodes can be either univariate transcendental functions or basic operations. $\nlcertify$ approximates $t$ with means of semialgebraic estimators and provides lower and upper bounds of $t$ over $\K$. When $t$ represents a polynomial, the tool computes lower and upper bounds of $t$ using a hierarchy of sparse SOS relaxations, as outlined in Section~\ref{sec:pop}. The extension to the semialgebraic case is straightforward through the implementation of the Lasserre-Putinar lifting-strategy~\cite{LasPut10lift}. The user can choose to approximate transcendental functions with maxplus estimators as well as best uniform (or minimax) polynomials.
The maxplus method derives lower (resp.~upper) estimators using concave maxima (resp.~convex infima) of quadratic forms (see Figure~\ref{fig:sampapprox_atn} for an example).
Alternatively, univariate minimax polynomials are provided with an interface to the $\sollya$ environment, in which the Remez iterative algorithm is implemented. In this way, $\nlcertify$ computes certified global estimators from approximations of primitive functions by induction over the syntax tree $t$.
\begin{example}[from LEMMA ${9922699028}$ Flyspeck \footnote{See the file available at~\url{http://code.google.com/p/flyspeck/source/browse/trunk/text_formalization/nonlinear/ineq.hl}}]
\label{ex:trans}
Let define the polynomial $\Delta \xb := x_1 x_4 ( - x_1 +  x_2 +  x_3  - x_4 +  x_5 +  x_6) 
 + x_2 x_5 (x_1  -  x_2 +  x_3 +  x_4  - x_5 +  x_6)  
 + x_3 x_6 (x_1 +  x_2  -  x_3 +  x_4 +  x_5  - x_6)  
 - x_2 x_3 x_4  -  x_1 x_3 x_5  -  x_1 x_2 x_6  - x_4 x_5 x_6$, the semialgebraic functions $r(\xb) := \partial_4 \Delta \xb / \sqrt{4 x_1 \Delta \xb}$ and $l(\xb) := 1.6294 - \pi / 2 - 0.2213 (\sqrt{x_2} +  \sqrt{x_3} +  \sqrt{x_5} +  \sqrt{x_6}  -  8.0) + 0.913 (\sqrt{x_4}  -  2.52) + 0.728 (\sqrt{x_1}  -  2.0)$, as well as the box $\K := [4, 2.1^2]^3 \times [2.65^2, 8] \times [4, 2.1^2]^2$. Note that for illustration purpose, the inequality has been modified by taking a sub-box of the original Flyspeck inequality box $[4, 2.52^2]^3 \times [2.52^2, 8] \times [4, 2.52^2]^2$.
 
Here we display and comment the output of $\nlcertify$\footnote{The parameter settings are \code{samp_iters = 3}, no branch and bound subdivisions (\code{bb = false}), \code{xconvert_variables = true},  \code{check_certif_coq = false}} for the inequality $\forall \xb \in \K, l(\xb) + \arctan (r(\xb)) \geq 0$.  The total (informal) computation time is about 20 seconds.
\begin{figure}[!h]
\begin{center}
\begin{tikzpicture}[scale = 0.67, level distance=15mm,
level 1/.style={sibling distance=60mm},
level 2/.style={sibling distance=20mm},
level 3/.style={sibling distance=30mm}]
\node[draw, thick, circle]{ $+$}
child {node[draw, thick, rectangle] {$l(\xb)$} }
child {node [draw, thick, circle]{$\arctan$}
child {node [draw, thick, rectangle]{$r(\xb)$}}};   	
\end{tikzpicture}
\caption{The abstract syntax tree of the function $f$ from LEMMA ${9922699028}$ Flyspeck}
\label{fig:ast_9922}
\end{center}
\end{figure}
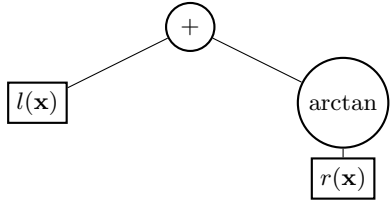
\vspace*{-1cm}
\lstset{language=NLcertify}
\scriptsize
\begin{lstlisting}
% ./nlcertify 9922699028_modified
Proving that - 1.5708 + atan ... >= 0 over the box  [(4, 4.41); 
(4, 4.41); (4, 4.41); (7.0225, 8); (4, 4.41); (4, 4.41)] ...
Bounding semialgebraic components
 Computing approximation of atan on [0.0297, 0.4165]
 Minimizer candidate x = [4; 4; 4; 8; 4; 4]
  Control points set: [0.3535] ...
Semialgebraic components bounded

Iteration 1
 Lower bound = -0.00463
 Minimizer candidate x = [4; 4; 4; 7.0225; 4; 4]
Iteration 2
  Control points set: [0.1729; 0.3535] ...
 Lower bound = -0.00006025
 Minimizer candidate x = [4; 4; 4; 7.6622; 4; 4]
Iteration 3
  Control points set: [0.1729; 0.2884; 0.3535] ...
 Lower bound = 0.000004662
 Minimizer candidate x = [4; 4; 4; 7.8083; 4; 4]
\end{lstlisting}
\normalsize
Lower and upper bounds for the semialgebraic components (i.e. $r$ and $l$) are computed using SOS relaxations. An interval enclosure for $r$ is $[m, M]$, with $m := 0.0297$ and $M := 0.4165$. Multiple evaluations of $f$ return a set of values and we obtain a first minimizer guess $\xopt{1}{4}{4}{4}{8}{4}{4}$ of $f$ over $\K$, which corresponds to the minimal value of the set.
Then, the solver performs three iterations of the nonlinear maxplus algorithm.
\begin{enumerate}[noitemsep,topsep=0pt,label={(\arabic*)}]
\setcounter{enumi}{0} 
\item The tool returns an underestimator $\parab_{a_1}$ of $\arctan$ over $[m, M]$, with $a_1 := r (\xb_1) = 0.3535$. Then, it computes
  $ m_1 \leq \min_{\xb \in \K} \{l(\xb) + \parab_{a_1} (r (\xb)) \}$.
   It yields $m_1 = -4.63 \times 10^{-3} < 0$ and $\xopt{2}{4}{4}{4}{7.0225}{4}{4}$. 
  
\item From the second control point, we get $a_2 := r (\xb_2) = 0.1729$ and a tighter bound
  $ m_2 \leq \min_{\xb \in \K} \{l(\xb) + \max_{1 \leq i \leq 2 } \{ \parab_{a_i} (r (\xb))\} \}$. 
We get $m_2 = -6.025 \times 10^{-5} < 0$ and  $\xopt{3}{4}{4}{4}{7.6622}{4}{4}$.

\item From the third control point, we get $a_3 := r (\xb_3) = 0.2884$ and
  $ m_3 \leq \min_{\xb \in \K} \{l(\xb) + \max_{1 \leq i \leq 3} \{ \parab_{a_i} (r (\xb))\} \}$. We obtain $m_3 = 4.662 \times 10^{-6} > 0$. Thus, the inequality is solved.
\end{enumerate}
\vspace*{-0.3cm}
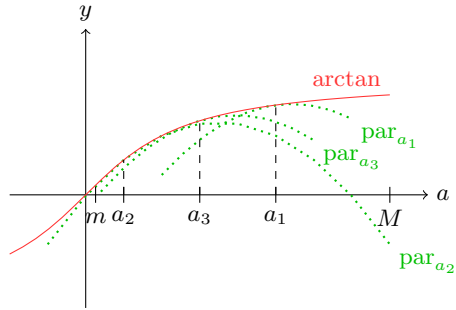
\begin{figure}[!htbp]
\begin{center}
\begin{tikzpicture}[scale=1]	
		
\draw[->] (-1,0) -- (4.5,0) node[right] {$a$};
\draw[->] (0,-1.5) -- (0,2.2) node[above] {$y$};

\draw[color=whitegreen, dotted, thick, samples=\samplepar] plot [domain=1:3.5]  (\x, {-0.32*(\x -\afst)^2 + 1/(1 + (\afst)^2)* (\x -\afst) + rad(atan(\afst)) }) node[anchor = north west] {$\parab_{a_1}$};
\draw[color=whitegreen, dotted, thick, samples=\samplepar] plot [domain=-0.5:4]  (\x, {-0.32*(\x -\asnd)^2 + 1/(1 + (\asnd)^2)* (\x -\asnd) + rad(atan(\asnd)) }) node[anchor = north west] {$\parab_{a_2}$};
\draw[color=whitegreen, dotted, thick, samples=\samplepar] plot [domain=0.2:3]  (\x, {-0.32*(\x -\athd)^2 + 1/(1 + (\athd)^2)* (\x -\athd) +  rad(atan(\athd))}) node[anchor = north west] {$\parab_{a_3}$};

\draw[color=red!80, samples=\sampleatn] plot [domain=-1:4] (\x,{rad(atan(\x))}) node[anchor = south east] {$\arctan$};
\draw (\lo,-0.3) node {$m$}; \draw (\up,-0.3) node {$M$}; \draw (\up,-3pt) -- (\up,3pt); \draw (\lo,3pt) -- (\lo,-3pt);
\draw (\athd,-0.3) node {$a_3$};  \draw (\athd ,3pt) -- (\athd ,-3pt); \draw [black, dashed] (\athd,0) -- (\athd,{rad(atan(\athd))});
\draw (\asnd,-0.3) node {$a_2$};  \draw (\asnd ,3pt) -- (\asnd ,-3pt); \draw [black, dashed] (\asnd,0) -- (\asnd,{rad(atan(\asnd))});
\draw (\afst,-0.3) node {$a_1$};  \draw (\afst ,3pt) -- (\afst ,-3pt); \draw [black, dashed] (\afst,0) -- (\afst,{rad(atan(\afst))});
\end{tikzpicture}
\caption{A hierarchy of maxplus quadratic underestimators for $\arctan$}	\label{fig:sampapprox_atn}
\end{center}
\end{figure}
\end{example}		
\vspace*{-1.3cm}
\section{Conclusion}
$\nlcertify$ aims at combining the safety of the $\coq$ proof assistant with the efficiency of informal optimization algorithms, based on low degree maxplus estimators and sparse semialgebraic relaxations.  This could allow to derive safe solutions for challenging problems that require to certify both approximation of transcendental functions and bounds for polynomial programs such as impulsive Rendezvous problems.
Further developments on the formal side include the integration of a new reflexive tactic inside the $\coq$ standard library.  Adding faster arithmetic for the polynomial coefficients ring would speedup the computation of the SOS checker. The current features could also be extended to handle noncommutative SOS certificates as well as discrete combinatorial optimization.
\if{ Certif of other problems: 
1) Program verification in static analysis $x_{k+1} = f (x_k)$
}\fi


\end{document}